# Spin excitations in the quasi-two-dimensional charge-ordered insulator $\alpha$-(BEDT-TTF)$_2$I$_3$ probed via $^{13}$C NMR

Kyohei Ishikawa[1], Michihiro Hirata[1,†,*], Dong Liu[1,*], Kazuya Miyagawa[1],

Masafumi Tamura[2], and Kazushi Kanoda[1,*]

[1]*Department of Applied Physics, University of Tokyo, Bunkyo-ku, Tokyo 113-8656, Japan*
[2]*Department of Physics, Faculty of Science and Technology, Tokyo University of Science, Noda, Chiba 278-8510, Japan*
[†]*Current address: Institute for Materials Research, Tohoku University, Aoba-ku, Sendai 980-8577, Japan*

**ABSTRACT**

The spin excitations from the nonmagnetic charge-ordered insulating state of $\alpha$-(BEDT-TTF)$_2$I$_3$ at ambient pressure have been investigated by probing the static and low-frequency dynamic spin susceptibilities via site-selective nuclear magnetic resonance at $^{13}$C sites. The site-dependent values of the shift and the spin-lattice relaxation rate $1/T_1$ below the charge-ordering transition temperature ($T_{\mathrm{CO}} \approx 135$ K) demonstrate a spin density imbalance in the unit cell, in accord with the charge-density ratio reported earlier. The shift and $1/T_1$ show activated temperature dependence with a static (shift) gap $\Delta_{\mathrm{S}} \approx 47$–$52$ meV and a dynamic ($1/T_1$) gap $\Delta_{\mathrm{R}} \approx 40$ meV. The sizes of the gaps are well described in terms of a localized spin model, where spin-1/2 antiferromagnetic dimer chains are weakly coupled with each other.

## I. INTRODUCTION

The organic layered salt $\alpha$-(BEDT-TTF)$_2$I$_3$ (abbreviated as $\alpha$-I$_3$ hereafter) is composed of alternately stacked conducting layers of (BEDT-TTF)$^{+1/2}$ molecules and nonmagnetic insulating layers of triiodide anions (I$_3$)$^{-1}$ [1], where BEDT-TTF (or ET) stands for bis(ethylenedithio)-tetrathiafulvalene. In the conducting layers, a quasi-two-dimensional (quasi-2D) electronic system possessing a 3/4-filled energy band is realized in which the unit cell contains four ETs with three nonequivalent molecular sites at room temperature ($T$), distinguished as $A$ (= $A'$), $B$, and $C$ [Fig. 1(a)]. The electronic states around the Fermi energy $E_F$ are described by the molecular orbitals associated to these molecular sites in the unit cell, as is generally the case for ET-based organic compounds [2].

Under a pressure ($P$) above 12 kbar, it has been shown that 2D Dirac cones appear in the conducting layers which have gapless points fixed at $E_F$ [3-6]. The low-energy excitations around the gapless points are described by massless Dirac fermions (DFs) with anisotropic linear energy-momentum dispersion [7]. At ambient $P$, the Dirac cones are likely present at high $T$, since recent $^{13}$C-NMR studies found, in the local spin density $\sigma_i$, a large difference among nonequivalent sites $i$ (= $A$, $A'$, $B$, and $C$) − $\sigma_B < \sigma_A$ (= $\sigma_{A'}$) $< \sigma_C$ − with strong $T$ dependence. It has been revealed that this feature is characteristic for the high-energy part of the DF-type excitations in a tilted Dirac-cone system like $\alpha$-I$_3$ [8-12]. However, the presence of the low-energy linear spectrum near $E_F$ is hidden at low $T$ beneath the first-order charge-ordering (CO) transition taking place at $T_{CO} \approx 135$K [1,3,13-27] due to strong electron correlations [28,29]. The transition is accompanied by an opening of an energy gap in the charge and spin excitation spectra and a formation of quasi-one-dimensional (quasi-1D) charge stripes along the crystalline $b$ axis [14], leading to an insulating spin-singlet ground state [1,3,17,18,20,30,31]. The inversion centers, locating on the molecules $B$ and $C$ and in between the molecules $A$ and $A'$ at $T > T_{CO}$ [Fig. 1(a)], vanish below $T_{CO}$ such that all of the four sites in the unit cell become nonequivalent [Fig. 1(b)] [14,27,32]. The ratio of the localized ($i$th-site) charge $\rho_i$ for the charge-rich sites ($\rho_A \approx \rho_B$) to the charge-poor sites ($\rho_A \approx \rho_C$) is estimated to be 3–4, according to x-ray [14], nuclear magnetic resonance (NMR) [32], infrared absorption [26], and Raman experiments [15,16].

In the CO state of $\alpha$-I$_3$, the low-energy electronic properties have been investigated with a particular emphasis upon the understanding of the charge excitations. Optical measurements revealed a charge gap of 75 meV and detected an in-gap tail in the excitation spectrum under a $b$-axis polarized light [33]. The dc transport measurements revealed a range of charge gaps ($2\Delta^{charge} \approx 40$–120 meV), which have a large anisotropy in the crystalline $ab$ plane [17,18,34]. Uji et al. [17] proposed a Kosterlitz-Thouless-like transition based on an excitonic model and argued that the variation of $\Delta^{charge}$ can be accounted for by the sample qualities. Ivek et al. [33,34] investigated the dielectric response and found long wavelength excitations with an in-plane anisotropic dispersion as well as short wavelength excitations which were discussed in terms of the motion of domain-wall pairs. Additionally, collective charge responses reminiscent of sliding charge density waves were suggested [35]. In contrast to the charge sector, where the static and dynamic properties of the CO state are fairly well understood, the spin sector has not yet been investigated in detail, and little is known about its nature. In this paper, we investigated the spin excitations in the CO state of α-I$_3$ at ambient $P$ by measuring the static susceptibility $\chi$ (the Knight shift $K$) and the low-frequency dynamic susceptibility (the nuclear spin-lattice relaxation rate $1/T_1$) through $^{13}$C-NMR measurements in the $T$ range from 60 to 200 K. We evaluated the static and dynamic spin gaps as well as the spin density profile in the CO state and discussed the results in light of their origins.

## II.  EXPERIMENTAL

Single crystals of α-I$_3$ were prepared by the conventional electrochemical oxidization method. Nuclear magnetic resonance measurements were performed in an external magnetic field ($H$) of 6 T for $H \| ab$, where the in-plane field direction, specified by the angle $\psi$, was measured from the crystalline $a$ axis in the conducting $ab$ plane [see the inset of Fig. 1(a)]. For obtaining $^{13}$C-NMR signals, 99% of the central double-bonded carbon atoms in ET molecules were selectively enriched by $^{13}$C-isotopes with a nuclear spin $I = 1/2$ [inset of Fig. 1(b)]. Here, $^{13}$C-NMR spectra were obtained by the fast Fourier transformation of the spin-echo signals recorded at a fixed frequency using a commercially available spectrometer. For the origin of the NMR shift, the $^{13}$C resonance frequency of tetramethylsilane, (CH$_3$)$_4$Si (TMS), was used. The $^{13}$C spin-lattice relaxation time $T_1$ was obtained by an exponential fit to the recovery curve of the nuclear magnetization after saturation. Note that the difference in the NMR shift and $1/T_1$ between the two $^{13}$C nuclei at the molecular center [inset of Fig. 1(b)] does not matter in this paper, and thus, we shall not distinguish them hereafter. Namely, the value of the NMR shift and $1/T_1$, measured at the four molecular sites ($A$, $A'$, $B$, and $C$ in Fig. 1), refers to an averaged value for the two central $^{13}$C nuclear sites in the corresponding ET molecule. To characterize the charge gap, we measured the dc resistivity for 10 different single crystals by standard four-terminal methods.

## III.  RESULTS

### A. Static spin gap in the CO state

Figure 2(a) shows the $T$ dependence of the NMR spectra measured for $H \| ab$ [$\psi = 110°$; for the definition, see the inset of Fig. 1(a)]. Sharp NMR spectra split by the dipolar interaction were observed above $T_{CO} \approx 135$ K, reflecting the four molecular sites in the unit cell [8,10]. The line positions exhibit large angular dependence on varying the field direction $\psi$ [Fig. 3(a)], which can be employed to assign the spectra into the dipole split two doublets from the sites $B$ and $C$ and one quartet from the sites $A$ and $A'$ (= $A$) [8]. We determine the NMR shift (from TMS) for the sites $i$ (= $A$, $A'$, $B$, and $C$), $S_i$, from the center of gravity of the doublet or the quartet, which is given as $S_i(T, \psi) = K_i(T, \psi) + \delta_i(\psi)$, where $K_i$ is the electron spin shift (the Knight shift) and $\delta_i$ is the $T$-independent core-electron contribution (the chemical shift) at the corresponding site $i$. The $T$ dependence of the spectra exhibits a sudden change at $T \approx T_{CO}$ [13,32], where the $A$-site spectra split due to the inversion symmetry breaking, which makes the sites $A$ and $A'$ nonequivalent [Fig. 1(b)] [1,14,27,32]. The modulus of the NMR shift at all sites shows a discontinuous change at $T_{CO}$, and its $T$ dependence levels off below $\approx$ 70 K [Fig. 2(b)], indicating the vanishing static susceptibility at low $T$ owing to the spin gap opening.

To extract the spin part from the total shift, the NMR shift $S$ (for $\psi$ = 110°) is compared to the bulk magnetic susceptibility $\chi^{bulk}$ reported by Rothaemel *et al.* [31] as shown in Fig. 2(c), where the low-$T$ Curie-Weiss component is subtracted from $\chi^{bulk}$. A liner relation is observed between $S$ and $\chi^{bulk}$ for the sites $B$ and $C$, which guaranties that $S$ probes the static susceptibility. From the y-intercept of this (so-called $K$-$\chi$) plot, we obtain the chemical shift $\delta(110°)$ = 113 and 85 ppm for the sites $B$ and $C$, respectively. By subtracting $\delta$ from $S$, one obtains the $T$ dependence of the Knight shift at the site $i$.

In quasi-1D gapped spin systems, the simplest and most likely ground state is a dimer singlet state in which the low-lying spin excitation is the singlet-triplet excitation. For this case, the $T$ dependence of the susceptibility should obey the form [36] $K \propto \exp(-\Delta_S/k_B T)/T$ at low $T$ ($\Delta_S$: the static spin gap). Since the CO state in $\alpha$-I$_3$ has a stripelike (quasi-1D) charge pattern [14], one of the tentative but rather realistic approaches is to fit the susceptibility based on this expression. As presented in Fig. 2(d), Arrhenius plots of $KT$ for the sites $B$ and $C$ show good agreements with the activation formula, showing the activation energy of $\Delta_S \approx$ 52 (47) meV at the $B$ ($C$) site. As for the sites $A$ and $A'$, $K$ is too small to extract the static gap owing to the small hyperfine-coupling constant at $\psi$ = 110°. We note that another widely used expression for the $T$ dependence of the shift derived for the spin-1/2 two-leg Heisenberg ladder systems is given by $K \propto \exp(-\Delta_S/k_B T)/T^{1/2}$ after Troyer *et al.* [36]. Figure 2(e) shows the Arrhenius fits to the data using this expression, which leads to the spin gap of $\approx$ 48 (44) meV at the $B$ ($C$) site. Since the sizes of the gaps are comparable in two models, the former expression, which is more general than the latter, shall be used in the following arguments of the spin gaps, as we will describe below.

**B. Electron spin density in the CO state**

Figures 3(a)-3(c) show the angle $\psi$ dependence of the NMR spectra at $T = 200$, 130, and 70 K, and Figs. 3(d)-3(f) present the $\psi$ dependence of the $i$th-site total shift $S_i(\psi)$ at the corresponding $T$ deduced from Figs. 3(a)-3(c). While $S_i(\psi)$ at $T = 200$ K shows large angular dependence at all sites [Fig. 3(d)], a distinct change appears in the $\psi$ dependence of $S_i(\psi)$ at 130 K ($< T_{CO}$) [Fig. 3(e)], where $S_{A'}(\psi)$ and $S_C(\psi)$ vary moderately with changing $\psi$, whereas $S_A(\psi)$ and $S_B(\psi)$ keep showing large angular dependence akin to the behaviors at $T > T_{CO}$. The distinct change observed at $T \approx T_{CO}$ can be associated to the spin gap opening [30,31] that simultaneously occurs with the stripe-type CO transition in this compound [14]. At $T = 70$ K, the Knight shift is supposed to vanish ($K_i \approx 0$) as the temperature dependence of the total shift $S_i(T)$ becomes negligibly small below this $T$ [Fig. 2(b)]. Thus, the observed angular dependence of $S_i(\psi)$ at $T \leq 70$ K can be attributed to the $\psi$ dependence of the chemical shift $\delta_i(\psi)$. Indeed, the calculated curves based on the chemical-shift tensors and the x-ray structural data at a similar $T$ [14,32] show good agreements with the observed angular dependence of $S_i(\psi)$ at all sites [solid curves in Fig. 3(f)]. Therefore, we assume that the fitting curves to the data of $S_i(\psi, T = 70$ K$)$ correspond to the chemical shift $\delta_i(\psi)$ and, by subtracting this term from the total shift, obtain the Knight shift $K_i(T, \psi)$ for each $T$ and angle $\psi$: $K_i(T, \psi) = S_i(T, \psi) - S_i(T = 70$ K$, \psi)$.

The $i$th-site Knight shift in this compound can be expressed as $K_i(T, \psi) = \bar{a}_i(\psi)\chi_i(T)$ [8], where $\chi_i(T)$ is the static susceptibility on the site $i$, which is independent of $\psi$ [30], and $\bar{a}_i(\psi)$ is the $T$-independent hyperfine-coupling constant. It is evident that the anisotropy of the hyperfine-coupling constant is huge when one varies the field direction in the crystalline $ab$ plane [8], as reflected in the large $\psi$ dependence of the total shift at $T > T_{CO}$ [Fig. 3(d)]. We made a sinusoidal fit to the $\psi$ dependence of $K_i(T, \psi)$ and determined the in-plane isotropic ($K_i^{iso}$) and anisotropic ($K_i^{aniso}$) parts of the $i$th-site Knight shift using the expression

$$K_i(T,\psi) = K_i^{iso}(T) + K_i^{aniso}(T)\sin[2(\psi + \Phi_i)], \qquad (1)$$

where $\Phi_i$ is a $T$-independent phase which is determined by the crystallographic arrangement of the ET molecules in the $ab$ plane and is fixed to the value determined at 200 K ($\Phi_{A,A'} = 13.6°$, $\Phi_B = -111.3°$ and $\Phi_C = -93.0°$), and both $K_i^{iso}$ and $K_i^{aniso}$ are proportional to $\chi_i$ [8]. Since the magnitude of the susceptibility is small in the CO state, it is very important to convert $K_i$ into $\chi_i$ in a situation where the hyperfine-coupling constant is large in order to minimize the error bar. For this, we use $K_i^{aniso}$ ($> K_i^{iso}$) together with the anisotropic part of the in-plane hyperfine-coupling constant $\bar{a}^{aniso} = 7.2$ kOe/$\mu_B$ [8] that is $T$ and *site* independent in the conducting state. Note that $K_i^{iso}/K_i^{aniso}$ varies little (~ 5 % at most) between 140 K ($> T_{CO}$) and 130 K ($< T_{CO}$) at the charge-rich sites $A$ and $B$, suggesting that the variation of $\bar{a}^{aniso}$ across $T_{CO}$ can be practically neglected. Thus, we evaluated $\chi_i(T)$ from the relation $\chi_i(T) = K_i^{aniso}(T)/\bar{a}^{aniso}$ at all $T$ for all sites $i$.

In Fig. 4(a), we show the $T$ dependence of $\chi_i(T)$ at the sites $i = A, A', B$, and $C$. In the CO state right below $T_{CO}$, a marked difference in the sizes of the site-specific spin density $\sigma_i$ ($\propto \chi_i$) is observed which reaches $\sigma_B$ ($\approx \sigma_A$):$\sigma_C$ ($\approx \sigma_{A'}$) ~ 3:1 at $T = 130$ K. The observed spin density ratio (of $\sigma_B$ to $\sigma_C$) agrees well with the corresponding charge-density ratio $\rho_B/\rho_C$ [14-16,32] and is consistent with the commensurate charge stripe order, where the charge rich ($\rho_B \approx \rho_A$) and poor ($\rho_C \approx \rho_{A'}$) strips alternately order along the crystalline *a* axis [Fig. 4(c)] [14]. Note that the spin density ratio $\sigma_B/\sigma_C$ shows distinct characters in the CO state ($\sigma_B/\sigma_C > 1$) and in the conducting state ($\sigma_B/\sigma_C < 1$), in contrast to the charge-density ratio that remains $\rho_B/\rho_C > 1$ at all $T$, as shown in Fig. 4(b) [14]. The sudden change of $\sigma_B/\sigma_C$ at $T_{CO}$ can be understood by a discontinuous change of the electronic band structure across the first-order CO transition [21]. In the conducting state ($T > T_{CO}$), it has been revealed that the spin density is unrelated to the charge density [4,37]. Namely, the former is determined by the local density of states around $E_F$ within a window of $k_B T$, whereas the latter is affected by the global character of the band structure with a bandwidth of subelectronvolts [4]. In the CO state ($T < T_{CO}$), however, the conducting electrons are localized on the ET sites. The size of the spin density $\sigma_i$ should then be proportional to the amount of the charge density $\rho_i$ at the corresponding site, as is indeed the case here.

### C. Low-frequency spin excitations and dynamic spin gap in the CO state

Figure 5(a) shows the $T$ dependence of $(1/T_1)_i$ ($i = A, A', B$, and $C$) measured for $H \parallel ab$ ($\psi = 110°$). At $T > T_{CO}$, $1/T_1$ shows moderate $T$ dependence at all sites in accordance with the earlier results [8,10]. With decreasing $T$ across $T_{CO}$, $1/T_1$ for the spin-rich sites ($A$ and $B$) shows an abrupt enhancement at $\approx T_{CO}$, whereas a discontinuous drop of $1/T_1$ is observed for the spin-poor sites ($A'$ and $C$) at the same $T$. On further cooling, an exponential decrease of $1/T_1$ develops at all sites, indicating the spin gap opening in agreement with the result of $\chi_i$ [Fig. 4(a)]. However, the observed sudden increase of $(1/T_1)_i$ ($i = A, B$) at $T_{CO}$ is in clear contrast to $\chi_i$ ($i = A, B$), which drops immediately after the transition [Fig. 4(a)]. The distinct $T$ dependence of $\chi_i$ and $(1/T_1)_i$ right at $T_{CO}$ can be accounted for by the emergent difference in the spin density between rich and poor sites that abruptly grows at the onset of the CO transition from the itinerant state to the localized state. The relaxation rate is proportional to the square of both the hyperfine-coupling constant and the spin density. For the current field orientation ($\psi = 110°$), the hyperfine-coupling constants for $B$ and $C$ sites are nearly the same [8], indicating that the observed enhancement of $1/T_1$ can be ascribed to the discontinuous change of the spin densities at $T_{CO}$. In fact, at 130 K ($< T_{CO}$), the ratio of the square root of the relaxation rate reads $[(1/T_1)_B/(1/T_1)_C]^{1/2} \sim 2.8$, which agrees well with the rich/poor ratio of the spin density, $\sigma_B/\sigma_C \sim 3$ (extracted from $\chi$), suggesting that the relation $(1/T_1)^{1/2} \propto \chi$ holds. The overall agreements between the values of $\rho_B/\rho_C$, $\sigma_B/\sigma_C$ and $[(1/T_1)_B/(1/T_1)_C]^{1/2}$ are consistent with the picture of localized electrons within the unit cell, where the site-dependent local spin densities are essentially proportional to the corresponding charge densities.

Figure 5(b) shows the Arrhenius plot of $1/T_1$ for the spin-rich ($B$) and spin-poor ($C$) sites. We fitted the $T$ dependence of $1/T_1$ in a low temperature range with an activation form, $1/T_1 \propto \exp(-\Delta_R/k_BT)$ ($\Delta_R$: the dynamic spin gap) and obtained $\Delta_R \approx 40$ meV for the sites $B$ and $C$, which is to be compared to the static gap determined by the Knight shift $\Delta_S \approx 52$ (47) meV for the site $B$ ($C$). The difference between the static and dynamic gaps, although not sizable, shall be discussed below (in Sec. IV).

The relaxation rate $1/T_1$ at the $A$ and $A'$ sites becomes almost identical at $T < 80$ K despite the difference in the spin densities [Fig. 4(a)], indicating that there is an averaging effect of $T_1$ between these sites, which is in sharp contrast to the case of the $B$ and $C$ sites as seen in Fig. 5(a). This averaging effect is likely induced by the spin diffusion ensured by the nuclear spin-spin coupling [38] because the NMR lines show a large overlap for the sites $A$ and $A'$ [as one can see in Figs. 2(a) and 3(c)]. When $T_1$ exceeds the spin-spin relaxation time at low $T$, which appears to be the case here, an intersite averaging of $1/T_1$ can take place due to the spin diffusion. This prevents us from estimating the dynamic spin gap at the $A$ and $A'$ sites.

**IV. DISCUSSIONS**

To have a better understanding of the spin excitations in the CO state of $\alpha$-$I_3$, the spin gaps revealed in this paper are compared to the charge excitation gaps [18,20]. First, we recall that the optical conductivity measurement reveals an energy gap of 75 meV, whereas the values of the dc transport gap are rather distributed around the optical gap in the 40–120 meV range [17,18,34]. To reconcile this distribution of gap values, we have performed dc transport measurements using standard four-terminal methods and evaluated the charge gap in 10 different single crystals (dubbed #1–#10). Except for a few samples, the majority of samples provide similar results as summarized in Fig. 6, where the $T$ dependence of the normalized resistance $R(T)/R(300\text{ K})$ for the samples #1–#8 is plotted as a function of $1/T$. In the intermediated $T$ range (for $1/T \approx 0.01$–$0.02$ K$^{-1}$), all the observed $R$ approximately falls into a single curve that is reconcilable with the optical band gap of 75 meV mentioned above. On further cooling, $R$ tends to saturate, which may be attributable to the extrinsic surface roughness or damage that allows leakage current in a highly resistive state and/or intrinsic edge states recently proposed by Omori *et al.* [39]. Indeed, the $P$ dependence of the resistivity suggests the presence of the edge states [20]. Both of these effects depend on the sample quality as well as the electrode geometries attached to the sample surface, presumably causing the sample dependence of the data below $T_\text{CO}$ as shown in Fig. 6, and appear to be irrelevant to the charge excitations in the bulk of the CO state. Thus, we consider that the resistivity data do not contradict the bulk charge excitation gap of approximately 75 meV revealed by the optical measurements, and therefore, the observed NMR spin gaps will be compared to this value of the charge gap.

Note that the optical conductivity and low-frequency dielectric measurements found low-energy charge excitations which have an excitation-energy scale much smaller than the size of the spin gap [33]. Since such low-energy excitations were not captured by the present NMR measurements, the optical and dielectric excitations at low energy turn out to be spinless excitations. The activated phason-like modes and domain-wall motions described in Refs. [33,34] are compatible with this picture.

Now, we focus on the magnetic excitations in the CO state in more detail. The present NMR measurements reveal that the spin gaps, both static $\Delta_\text{S}$ ($\approx 47$–52 meV) and dynamic $\Delta_\text{R}$ ($\approx 40.0$ meV), are substantially smaller than the charge gap $2\Delta^\text{charge}$ ($\approx 75$ meV), as is often the case in correlation-induced insulators [40]. In these, the spin excitations have a smaller characteristic energy scale than what is expected for quasiparticle excitations. Furthermore, the discontinuous increase of $1/T_1$ at $T_{CO}$ observed on the sites $A$ and $B$ cannot be brought about for quasiparticle excitations in canonical band insulators, where $1/T_1$ drops immediately at the onset of the gap opening. In the CO state, we recall that the charges are localized on the ET lattice sites and form charge rich ($A$, $B$) and poor ($A$', $C$) strips along the crystalline $b$ axis. Since there is a correlation in the local charge and spin densities [Fig. 4(b)], the charge stripes can be simultaneously considered as quasi-1D spin-1/2 chains as shown in Fig. 4(c).

Figure 7(a) depicts the anisotropic network of the nearest neighbor hopping amplitudes between nonequivalent sites at $T < T_{CO}$, which is characterized by 12 transfer integrals [14] indicated by $a1$ to $b4$ in the figure. According to the density functional calculation [5], the largest hopping is present between the spin rich sites $A$ and $B$ ($b2$'), but the second largest ones [along the paths $A$-$C$ ($b1$') and $B$-$A$' ($b2$)] have similar sizes, consistent with the crystal structure determined by the synchrotron x-ray diffraction measurement [14]. In quasi-1D organic systems (TMTTF)$_2$X (TMTTF: tetramethil-tetrathiafulvalene, X: anion), it has been shown that similar nonuniform networks of transfer integrals are present among molecules and that the nearest neighbor Coulomb repulsions lead to a CO state, where a pair of charge-rich and charge-poor molecules accommodates one electron [41]. A similar situation is expected between the charge rich and poor sites in $\alpha$-I$_3$; namely, the sites $A$ and $C$ (and also $B$ and $A$') form a dimer and accommodate one electron. This simplification allows one to reduce Fig. 7(a) into a half-filled spin lattice model in which the $AC$ and $BA$' dimers can be regarded as new sites carrying spin-1/2 moments, as shown in Fig. 7(b). However, in contrast to (TMTTF)$_2$X where the spins are intact at the transition, the CO transition in $\alpha$-I$_3$ is accompanied by a concomitant spin-singlet formation as a consequence of the strong dimerization [1,3,17,18,20,30,31].

The spin moments on each dimer interact through an effective exchange coupling $J(t_{ij}^l) = 4(t_{ij}^l)^2 \langle n_i n_j \rangle_l / U$ ($i, j = A, A', B$, and $C$) [42], where $t_{ij}^l$ is the transfer integral between the molecules $i$ and $j$ along the path $l$ (abbreviated to be $t_l$ hereafter), $\langle n_i n_j \rangle_l$ is the nearest neighbor charge correlation function for the same path, and $U$ is the on-site Hubbard interaction. Since we do not have exact values of $\langle n_i n_j \rangle_l$ which reproduce the behaviors of the complex real system, we approximate the correlation function by the product of charge densities at the corresponding molecular sites $\rho_i \rho_j$ as a pragmatic approach, which reads $J(t_l) \sim 4(t_l)^2 \rho_i \rho_j / U$.

Then by assuming the values of $\rho_i$ and $t_l$ reported by the x-ray diffraction measurement [14] in conjunction with the empirical value of the Hubbard interaction ($U = 1$ eV [41]), the values of the exchange coupling are estimated as presented in Table I.

As seen in Table I and Fig. 7(b), the interdimer magnetic couplings along the crystalline $b$ axis provides the dominant contributions to the magnetic interactions, constituting a quasi-1D localized spin-1/2 chains along the $b$ axis. If one neglects all the interchain interactions together with the next nearest neighbor couplings as the first approximation, the dimers $AC$ and $BA'$ in each chain are connected via alternating transfer integrals ($b2'$ and $b3$), which can be mapped onto a $S = 1/2$ 1D Heisenberg antiferromagnetic (AF) model with staggered exchange couplings $\hat{H} = J \Sigma_{(\alpha,\beta)} S_\alpha \cdot S_\beta + J' \Sigma_{(\alpha,\beta)} S_\alpha \cdot S_\beta$, where $\alpha$ ($= AC$ and $BA'$) indicates the dimer site, $S_\alpha$ is the spin-1/2 operator at the site $\alpha$, $(\alpha,\beta)$ denotes nearest neighbor sites, and $J$ and $J'$ ($J > J' > 0$) are the AF exchange couplings.

For small but finite $J'$, Hida [43] has evaluated the triplet excitation energy from the singlet ground state for this model and revealed collective excitations with dispersion $E(k) = J - J'/2 \cos(2ka)$ ($k$: the wave vector, $a$: the intersite spacing). The spin gap is then given by the lowest excitation energy around the bottom of the triplet band (at $k = 0$), which is evaluated to be $\Delta_{model} \sim 71$ meV by using $J(t_{b2'}) = 76$ meV and $J(t_{b3}) = 11$ meV for the present case. Tanaka [44] has recently evaluated the charge correlation functions for the bonds $b2'$ and $b3$ in the ground state of $\alpha$-I$_3$ by variational calculations using the extended Hubbard model and revealed $<n_A n_B>_{b2'} = 0.54$ and $<n_A n_B>_{b3} = 0.66$. With these values, the exchange interactions are re-estimated as $J(t_{b2'}) = 69$ meV and $J(t_{b3}) = 12$ meV. Then one obtains the spin gap of $\sim 63$ meV, in reasonable agreement with the experimental values $\Delta_R \approx 40$ meV and $\Delta_S \approx 47$–52 meV, albeit assuming a simplified model.

As seen in Fig. 7(b), there are finite interchain couplings in the real system $J(t_{b1}) = 4.4$ meV, $J(t_{b3'}) = 3.8$ meV and $J(t_{a2}) = 2.8$ meV, which would cause 2D dispersion and should lower the gap. The observed difference in the size of the static gap $\Delta_S$ (measured by the NMR shift) and the dynamic gap $\Delta_R$ (measured by $1/T_1$) may reinforce this notion that the magnetic excitations are more 2D-like rather than 1D-like in the CO state of $\alpha$-I$_3$. Itoh and Yasuoka [45] have compared the ratio of these gaps $\Delta_R/\Delta_S$ for a wide range of $S = 1/2$ spin-gapped systems and revealed $\Delta_R/\Delta_S = 1$ for simple (1D) dimers while $\Delta_R/\Delta_S \neq 1$ for quasi-1D ladders and more complex 2D systems. The observed gap ratio in $\alpha$-I$_3$ is ~ 0.7–0.8, which does not much differ from unity as in the two-leg ladder system SrCu$_2$O$_3$ [46,47] and the Shastry-Sutherland system SrCu$_2$(BO$_3$)$_2$ [48] ($\Delta_R \leq 1.73\Delta_S$) yet is still different from the simple 1D dimer system CsV$_2$O$_5$ ($\Delta_R = \Delta_S$) [49]. This shows that the observed difference in $\Delta_R$ and $\Delta_S$ is rather compatible with the quasi-1D or 2D cases, suggesting the presence of finite interchain couplings. Gaining deeper knowledge on the nature of spin excitations in this 2D system naturally requires further theoretical studies by taking into account the realistic values for both of the on-site and nearest neighbor Coulomb repulsions, which are not explicitly considered here. It is an interesting future issue to see how the sizes of these gaps as well as the ratio $\Delta_R/\Delta_S$ vary on increasing $P$ and thereby suppressing $T_{CO}$ [3,19,20].

## V. CONCLUSIONS

The $^{13}$C-NMR shift and spin-lattice relaxation rate ($1/T_1$) have been measured together with the dc resistivity for the charge ordering system $\alpha$-I$_3$ at ambient pressure in the temperature range from 60 to 200 K. The site-specific spin density determined from the shift and $1/T_1$ values at 130 K yields a spin density ratio of ~ 3 between the spin-rich and spin-poor sites, which parallels the rich-poor ratio of the charge density suggested earlier in the stripelike charge-ordered state [14-16,26,32]. The observed good agreement between the spin-density and charge-density ratios together with the discontinuous jump in the value of $1/T_1$ right at the charge-ordering transition indicate a strongly localized spin character. The shift and relaxation rate $1/T_1$ exhibit activated temperature dependence below ≈ 135 K with the static (shift) gap of $\Delta_S \approx$ 47.4–51.7 meV and the dynamic ($1/T_1$) gap of $\Delta_R \approx$ 40.0 meV. The values of these gaps lie well below the charge gap of ≈ 75 meV, determined by optical measurements [31] and reconciled by the present dc transport measurements. The magnitudes of $\Delta_S$ and $\Delta_R$ and their difference ($\Delta_R/\Delta_S \approx$ 0.7–0.8) can be well accounted for by triplet excitations in a localized spin system accommodating a quasi-2D network of exchange interactions between dimeric $S = 1/2$ spins. The demonstration of the spin gap of 40 meV or larger suggests that the low-energy charge excitations captured by previous optical and dielectric measurements are spinless.

**ACKNOWLEDGMENTS**

The authors thank R. Kondo and H. Sawa for informing us about the structural data and the local electronic density distribution on triiodine anions before publication. We also thank N. Tajima, A. Kawamoto, and T. Takahashi for informing us about the detailed experimental results and Y. Suzumura, A. Kobayashi, H. Fukuyama, Y. Tanaka, and M. uOgata for fruitful discussions. This paper is supported by Ministry of Education, Culture, Sports, Science and Technology (MEXT) Grant-in-Aids for Scientific Research on Innovative Area (New Frontier of Materials Science Opened by Molecular Degrees of Freedom; no. 20110002), Japan Society for the Promotion of Science (JSPS) Postdoctoral Fellowships for Research Abroad (Grant No. 66, 2013), and JSPS Grant-in-Aids for Scientific Research (S) (no. 25220709).

K.I., M.H., and D.L. contributed equally to this work.

*E-mail: michihiro_hirata@imr.tohoku.ac.jp, dongliu.phy@gmail.com, kanoda@ap.t.u-tokyo.ac.jp

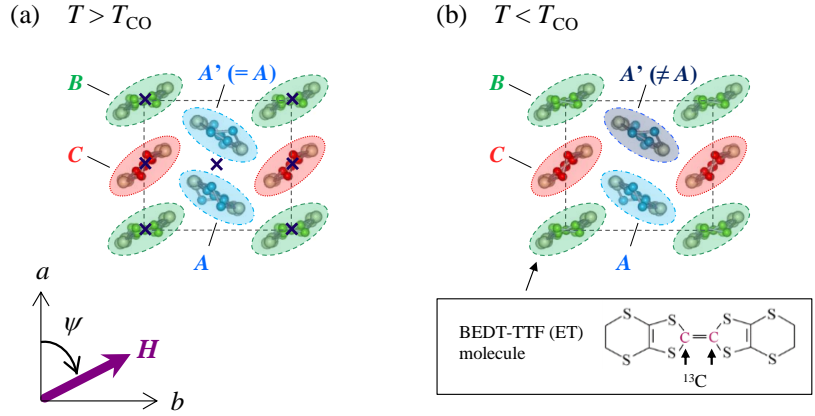

**FIG. 1**. Schematic illustrations of the crystal structures of $\alpha$-I$_3$ in the crystalline $ab$ plane (**a**) in the conducting state ($T > T_{CO}$) and (**b**) in the CO insulating state ($T < T_{CO}$) [14]. Nonequivalent ET molecules in the unit cell seen from the long axis of the molecule are indicated as $A$, $A'$, $B$, and $C$, and the crosses stand for the inversion centers which vanish in the CO state. Inset of (**a**): the definition of the field angle ($\psi$) of the applied magnetic field $H$ in the $ab$ plane, measured from the $a$ axis. Inset of (**b**): The molecular structure of the ET molecule. The positions of the $^{13}$C isotopes with a nuclear spin ($I = 1/2$) introduced for NMR measurements are indicated.

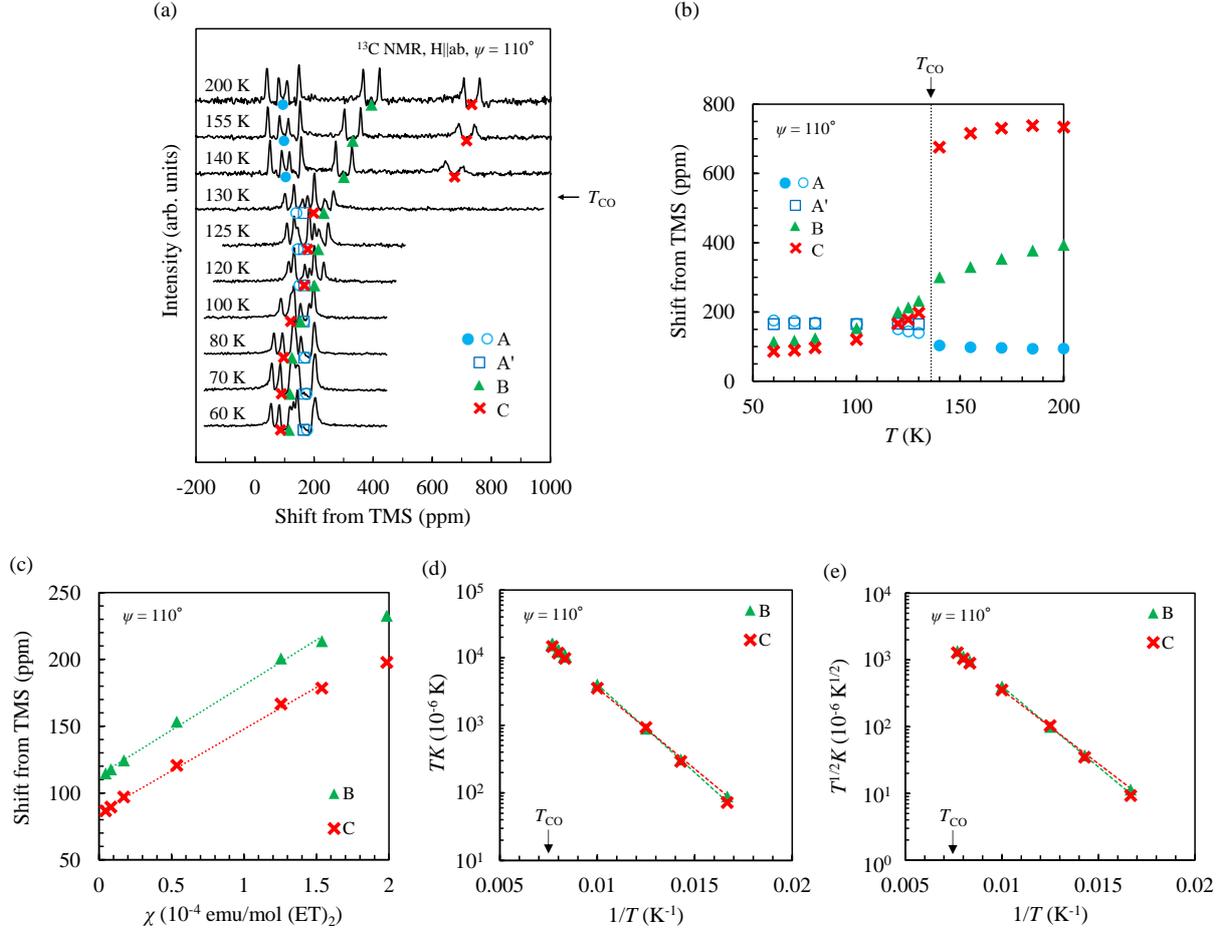

**FIG. 2.** Temperature dependence of (**a**) the $^{13}$C-NMR spectra and (**b**) the $i$th-site total shift $S_i$ for the sites $i = A$ (circles), $A'$ (squares), $B$ (triangles), and $C$ (crosses). $T_{CO}$ (≈ 135 K) indicates the CO transition temperature. The magnetic field of 6 T is applied in the crystalline $ab$ plane in the direction $\psi = 110°$ [see the inset of Fig. 1(a)]. The total shift $S_i$ in (b) indicates the center of gravity of the doublet (for the sites $B$ and $C$) or the quartet (for the sites $A$ and $A'$) in (a) (for details, see the text). (**c**) The total shift $S_i$ is plotted against $\chi$ for the sites $i = B$ and $C$, where $\chi$ is the bulk magnetic susceptibility given in Ref. [31], in which the low-$T$ Curie-Weiss component is subtracted. The $y$ intercept of this plot gives the value of the chemical shift at each site. (**d**) The Arrhenius plot of $TK_i$, which yields the static spin gap of $\Delta_S \approx 52$ and 47 meV for the sites $i = B$ and $C$, respectively. Note that the $i$th-site Knight shift $K_i$ is obtained from the total shift $S_i$ in (b) by subtracting the corresponding chemical shift value determined in (c). (**e**) The Arrhenius plot of $T^{1/2}K_i$, which yields the spin gap of $\Delta_S \approx 48$ and 44 meV for the sites $i = B$ and $C$, respectively.

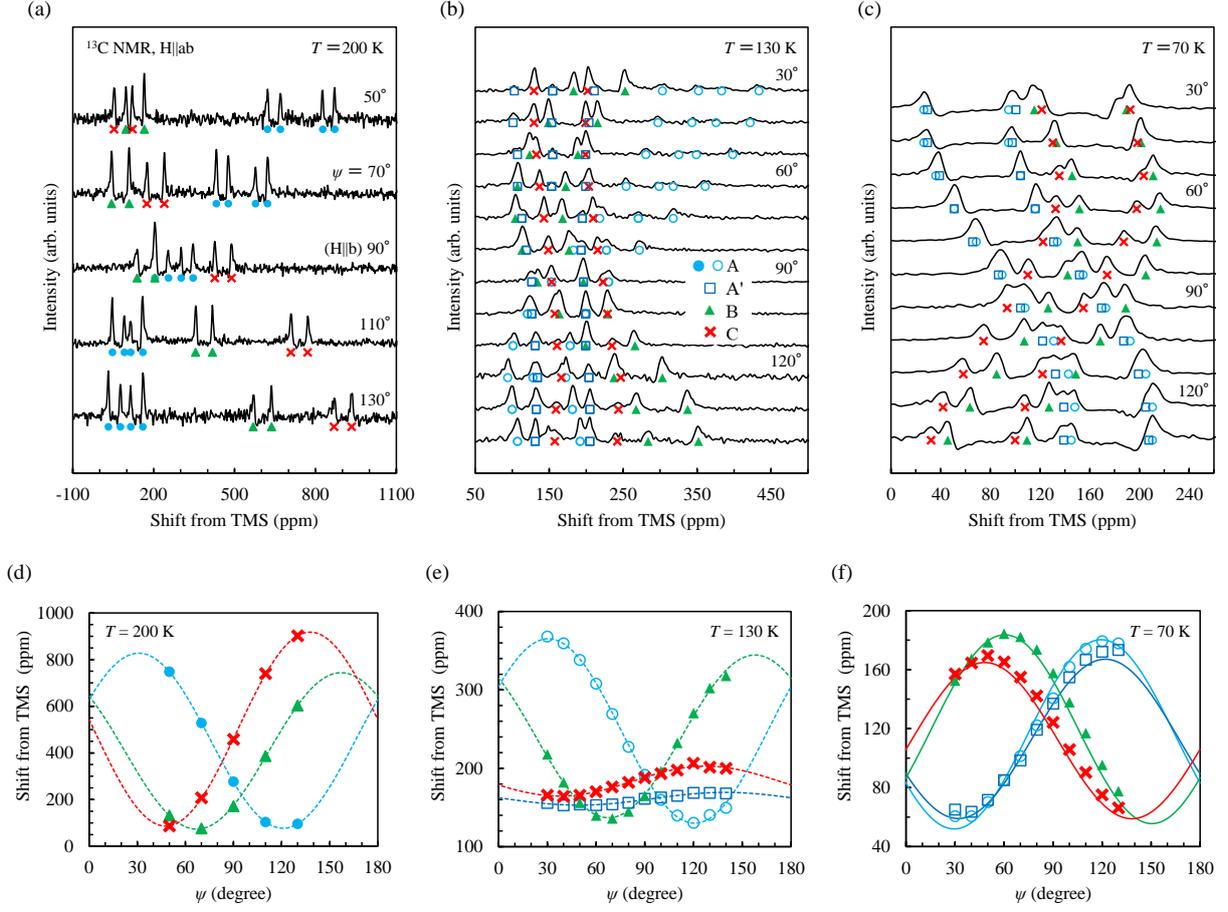

**FIG. 3.** The field orientation ($\psi$) dependence of **(a-c)** the $^{13}$C-NMR spectra and **(d-f)** the $i$th-site total shift $S_i(T, \psi)$ for the sites $i = A, A', B$, and $C$ measured at various $T$, where the magnetic field is rotated in the crystalline $ab$ plane. Same symbols used as in Fig. 2. The dashed curves in (d) and (e) represent the least-square sinusoidal fits to the data. The solid curves in (f) indicate the calculated $\psi$ dependence of the chemical shift for each site, which is obtained using the chemical-shift tensors reported in Ref. [32] and the x-ray crystal structure given in Ref. [14].

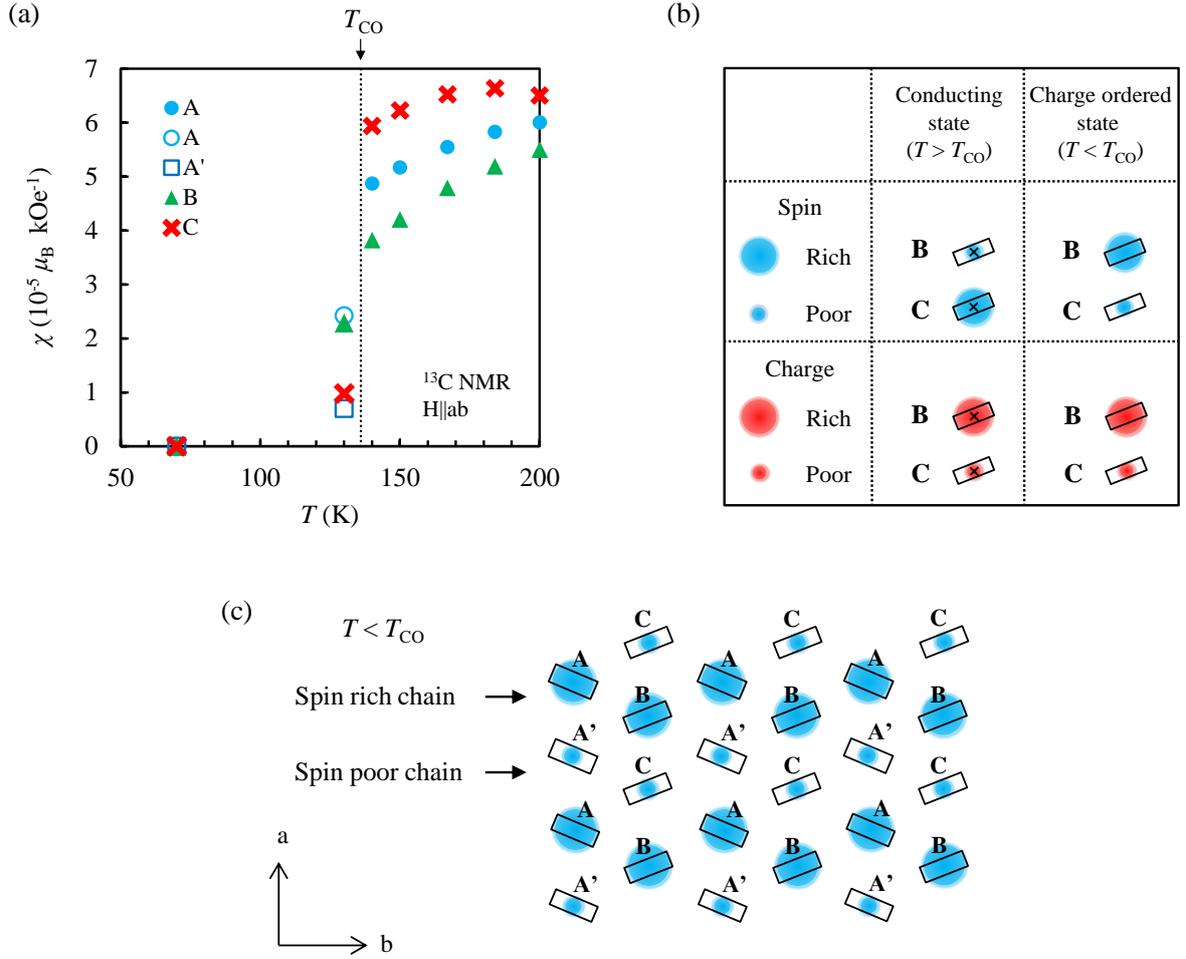

**FIG. 4.** (a) The temperature dependence of the site-specific electron spin susceptibility $\chi_i(T)$ for the sites $i = A, A', B$, and $C$. Same symbols used as in Fig. 2. The CO transition temperature $T_{CO}$ ($\approx$ 135 K) is indicated by the dotted vertical line. The susceptibility is obtained from the in-plane anisotropic part of the $i$th-site Knight shift $K_i^{aniso}(T)$, using the hyperfine-coupling constant $\bar{a}^{aniso} = 7.2$ kOe/$\mu_B$ [8] (see the text for details). (b) The schematic illustration of the spin density ($\sigma_i$), deduced from (a), and the charge density ($\rho_i$), reported in Ref. [14], at the sites $i = B$ and $C$ are shown for the conducting state ($T > T_{CO}$) and the CO insulating state ($T < T_{CO}$). Crosses stand for the inversion centers, which disappear at the onset of the CO transition. (c) The illustration of the spin density imbalance at $T < T_{CO}$ in the crystalline $ab$ plane deduced from the present NMR measurements.

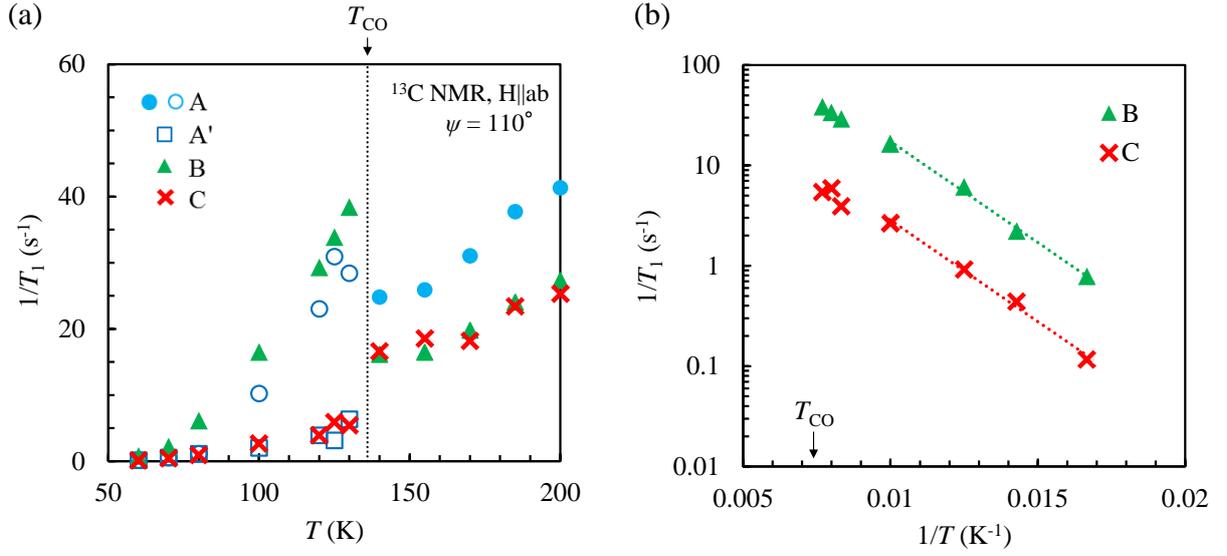

**FIG. 5. (a)** The temperature dependence of the $^{13}$C-nuclear spin lattice relaxation rate $(1/T_1)_i$ for the sites $i = A, A', B$, and $C$, obtained in the field orientation $\psi = 110°$ [same as in Fig. 2(a)]. Same symbols used as in Fig. 2, and the vertical dotted line indicates $T_{CO}$ ($\approx 135$ K). **(b)** The Arrhenius plot of $(1/T_1)_B$ and $(1/T_1)_C$ fitted at $T < 100$ K (dotted lines), which yields the dynamic spin gap of $\Delta_R \approx 40$ meV for the sites $i = B$ and $C$.

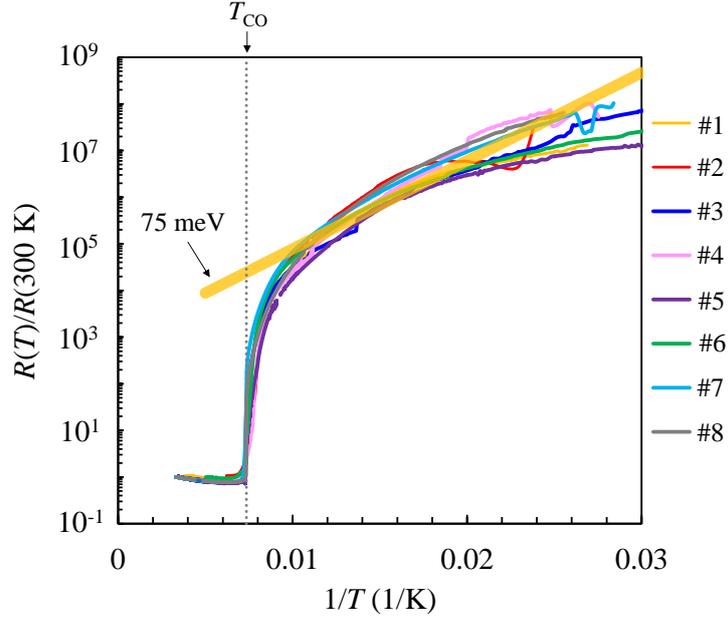

**FIG. 6.** The normalized electrical resistance in the crystalline *ab* plane of $\alpha$-$I_3$, plotted as a function of inverse temperature ($1/T$) for eight different samples, dubbed #1–#8. Bold straight line corresponds to the activation curve with the energy gap of 75 meV suggested by the optical measurement [33]. The dotted vertical line indicates $T_{CO}$ ($\approx$ 135 K).

(a) $T < T_{CO}$

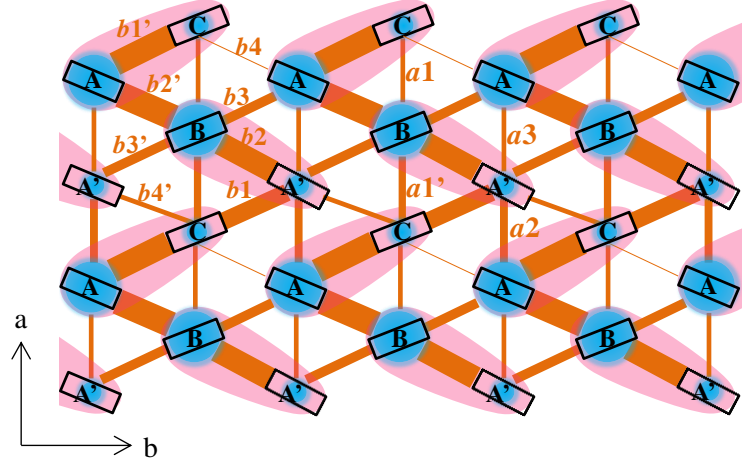

(b)

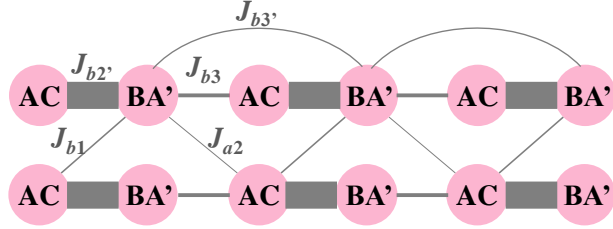

**FIG. 7.** (a) The 2D networks of nearest neighbor transfer integrals and the spin density imbalance [following Fig. 4(c)] in the crystalline *ab* plane at $T < T_{CO}$. The size of the filled circles reflects the modulus of the spin density on each site [as in Fig. 4(c)], while the thickness of the bonds connecting nearest neighbor sites indicates the magnitude of the transfer integrals $t_l$ for the bond $l$ (= $a1$–$b4'$), determined by the x-ray diffraction measurement [14]. Filled ellipses represent the spin densities that are spread due to the large hopping amplitudes $t_{b1'}$ and $t_{b2}$ connecting the spin-rich ($A$, $B$) and spin-poor ($A'$, $C$) sites (see the text for details). (b) An effective dimer spin model deduced from Fig. 7(a), where the two molecular pairs connected via $t_{b1'}$ ($A$ and $C$) and $t_{b2}$ ($B$ and $A'$) are regarded as dimers, dubbed $AC$ and $BA'$, respectively. Each circle indicates the dimer with a localized spin 1/2, which corresponds to the ellipses in (a). The spin-1/2 moments form antiferromagnetic Heisenberg chains with alternating exchange interactions in the horizontal direction, which are coupled to each other through interchain exchange interactions $J_{b1}$ and $J_{a2}$. The magnitudes of the effective exchange interaction are calculated by means of the expression $J(t_l) = 4t_l^2 \rho_i \rho_j / U$ [42,44] and are reflected as the thickness of the bonding lines in (b), where the values of the localized charge $\rho_i$ and the transfer integral $t_l$ (for the bond $l$) from Ref. [14] are used with the on-site Hubbard interaction of $U = 1$ eV [41] (see Table I).

| Bond $l$ | $(i, j)$ | $t_l$ (meV) | $\rho_i\rho_j$ | $J(t_{ij})$ (meV) |
|---|---|---|---|---|
| $b2'$ | $(A, B)$ | 178 | 0.60 | 75.86 |
| $b1'$ | $(A, C)$ | 165 | 0.21 | 23.22 |
| $b2$ | $(A', B)$ | 158 | 0.21 | 21.14 |
| $b3$ | $(A, B)$ | 67 | 0.60 | 10.75 |
| $b1$ | $(A', C)$ | 121 | 0.08 | 4.42 |
| $b3'$ | $(A', B)$ | 67 | 0.21 | 3.80 |
| $a2$ | $(A, A')$ | 54 | 0.24 | 2.77 |

**Table I**. List of the calculated effective exchange interactions (the right most column) $J(t_l) = 4t_l^2\rho_i\rho_j/U$ [42,44] for the bond $l$, connecting the nearest neighbor sites $(i, j)$, with the transfer integral $t_l$ and the local charge density $\rho_i$ at the sites $i$ (= $A$, $A'$, $B$, and $C$). The on-site Hubbard interaction $U$ of 1 eV [41] is assumed for the calculation. On the list, the values of the relevant transfer integrals (7 out of 12 from the top largest) together with the charge density at the sites $i$ ($\rho_i$) multiplied with that at the nearest neighbor sites $j$ ($\rho_j$), $\rho_i\rho_j$, are presented, which are determined by the x-ray diffraction measurement [14].